\documentclass[prb,twocolumn]{revtex4}
\begin{document}
\title{Magnetoresistance of a Wigner liquid in a parallel magnetic field}
\author{E.G.Batyev} \address{Institute of Semiconductor Physics,
Siberian Branch of Russian Academy of Sciences, Novosibirsk,
630090, Russia}\thanks{e-mail:batyev@isp.nsc.ru}
\begin{abstract} It is assumed that in a two-dimensional electron system
with strong correlation (a Wigner liquid), appearance of some
relatively slow-moving objects (clusters) composed of small number
of electrons is possible. Such clusters may exist in addition to
ordinary mobile carriers of the Fermi type. They can pin to
inhomogeneities and play the role of additional scatterers. The
clusters composed of two and three electrons are discussed (for
near order as in triangular lattice). The number of the clusters
depends on temperature and a parallel magnetic field (so
accordingly for conductivity and magnetization). In the frame of a
simple model, a resistivity increasing and a metal-insulator
transition with an increasing of a magnetic field are proved. Near
this transition, the resistivity changes with temperature
according to a linear law. The model gives a nonlinear dependence
of magnetization on a magnetic field.
\end{abstract}
\maketitle PACS: 71.27.+a

In recent years, low-density systems of two-dimensional electrons
(holes) are intensively studied (see, e.g., reviews \cite{1,2}).
That sort of systems exhibits a number of unusual properties. One
of the main properties is a metal-insulator transition which
occurs upon a decreasing of the carrier density. Moreover,  with
an increasing of a parallel magnetic field, a growth of
resistivity and a metal-insulator transition are observed. The
present paper is devoted to the theory of such systems in a
parallel magnetic field.

The above-mentioned phenomena may have various reasons. For
example, theoretically the effective mass $m^*$ could increase
with spin polarization, but the experiments apparently do not
manifest that. Further, interaction of carriers with impurities
can increase with spin polarization. This effect was considered in
the work \cite{3} where the interaction with charged impurities
was discussed. However, the concrete realization of the idea in
this work gives rise to doubts. The point is that, for the
screened interaction with charged impurity, the used in the work
\cite{3} formula is applicable only in the limit $r_S<<1$, while,
on the contrary, the limit $r_S>>1$ takes place for the considered
systems ($r_S$ is the dimensionless average distance between the
particles). In the work \cite{4}, another mechanism connected with
the presence of hole traps near the SiO$_2$/Si interface and their
possible recharging was discussed. Though this mechanism is
possible but its effect depends on a sample and changes from
specimen to specimen, whereas experiments show more or less
universal behavior for any utilized sample. Seemingly, it is
necessary to find some universal mechanism.

Apparently, the universal reason exists because of some universal
property of considered systems. Namely, a low-density system of
two-dimensional electrons (under $r_S>>1$) is the system with the
strong correlations, i.e. it has a near order in particle
arrangements (as in a Wigner crystal and, therefore, such a system
is called a Wigner liquid sometimes). The use of this property is
the main point of the present work.

First of all, it is necessary to do some remarks. In the limit
$r_S>>1$, there is the following hierarchy of energies. 1) The
Coulomb energy is the greatest one and is taken as the unity.
After all, it is this energy that results in a Wigner crystal and
gives a near order in a Wigner liquid. 2) The energy of zero
oscillations (or equivalently of typical plasma oscillations) is
of the order of $1/\sqrt{r_S}$. Apparently, because of the zero
oscillations the liquid do not crystallize right up to $r_S\approx
37$ (see the work \cite{5}). 3) At last, the Fermi energy is of
the order of $1/r_S$ if this energy is estimated by using of band
effective mass $m_b$ as in a Fermi gas. In reality, the Fermi
energy is even smaller because the effective mass $m^*$
renormalized due to interaction is greater than $m_b$. Further, we
shall deal with the energies of the order of the Fermi energy that
is smallest one among the all energies and is of the order of 
exchange energy (see below).

The main idea of the present work is connected with the conception
accepted in the work \cite{6}. In this work, an estimation of the
exchange interaction of particles (and, accordingly, of the
effective mass $m^*$) was done. For that, two neighboring
particles interacting with external potential produced by
environment and with each other via Coulomb were considered.
Either of the two takes up its position mainly in its potential
minimum, and from time to time they interchange their positions.
With the help of this model, the exchange splitting of energy
levels ($E_A-E_S$) (its dependence on $r_S$) has been found. The
exchange Hamiltonian that gives the splitting may be presented in
the form:
\begin{eqnarray}\nonumber H_{ex} =\lambda_S \biggl\{({\bf S_1
S_2})+ ({\bf Q_1 Q_2}) + \\ \label{1} 4({\bf S_1 S_2}) ({\bf Q_1
Q_2})-\frac{1}{4}\biggr\}\ ,\end{eqnarray} where the indices 1,2
are particle numbers, ${\bf S}$ is the spin operator, and ${\bf
Q}$ is the quasi-spin operator that corresponds to the two-valley
case ($Q=1/2$). Further, the two-valley case as at SiO$_2$/Si
inversion layer is examined. The constant (-1/4) in the brackets
is added for convenience. The value $\lambda_S$ and the effective
mass $m^*$ are connected with the energy splitting in the
following way: \begin{equation} \label{2} 2 \lambda_S=E_A-E_S \sim
\frac{n}{m^*}\ ,\end{equation} where $n$ is the carrier
concentration. The equation (\ref{2}) gives estimation of $m^*$.
Note that here the right-hand part is of the order of the Fermi
energy of quasi-particles.

{\bf Starting idea}. Further, one would think, it is necessary to
operate in accordance with the Landau theory of the Fermi liquid
(by using the effective mass estimation (\ref{2})). However, the
question arises about adequacy of similar approach in our case.
Actually, one may imagine a situation when, for example, a pair
energy is smaller than the double Fermi energy of quasi-particles,
and so a number of the pairs would arise together with the
quasi-particles.

In order to elucidate the situation, let us turn to the physical
picture which was implied in the work \cite{6}. In this work, the
exchange energy of two nearest neighbors was calculated. If take
into account all jumps of a particle in liquid, then one can see
that the minimal energy of a particle can be even lower than in a
pair, and this decrease is of the order of the exchange energy.
Obviously, this minimal energy can be attributed to quasi-particle
with zero momentum. So, as long as the all relevant to our problem
energies are of the same order, then, in particular, it is
permissible to compare a pair energy and the Fermi energy of
quasi-particles. Therefore, the above-mentioned effect can exist
and then influence of the pairs on the properties of the system
should be taken into account.

One can consider instead of the pairs some other formations. For
near order as in a Wigner crystal (i.e. as in triangular lattice),
a cluster composed of three nearest neighbors is the most
appropriate object. Obviously, energy of such a cluster (per a
particle) is smaller than energy of a pair, and so, probably, it
is necessary to do a preference for a three-particle cluster.

It is unlikely that one should consider more complicated objects
in liquid  because for that the ordered arrangement of particles
is necessary not only for nearest neighbors but for next nearest
neighbors as well. So further we shall consider the pairs and the
threes only.

Evidently, the number of the clusters depends on a magnetic field
(since clusters have spin) and temperature and the same relates to
the number of mobile carriers of the Fermi type (fermions). It is
natural to consider that the clusters are slow-moving formations
because their jumping amplitude is smaller than for a particle. It
means that they can be pinned even by small inhomogeneities and be
excluded from the electric current. Moreover, there is an
additional contribution to a resistance due to fermion scattering
by the pinned clusters. Therefore, the resistivity of the system
should depend on a magnetic field and temperature as well. The
above picture is a basis for proposed description of the
properties of the system.

In some extent, our approach is partly similar to the model used
in the work \cite{4}. However, unlike this work, we appeal
directly to the intrinsic properties of the strong correlated
system rather than to the external factors.

{\bf Clusters}. First of all, it is necessary to describe the
properties of the clusters. The states and energies of the
clusters are meant. Let $S$ and $Q$ be the total spin and
quasi-spin of a cluster, accordingly. We start with a pair. The
quantum numbers corresponding to its minimal energy are $S=1,\
Q=0$ or $S=0,\ Q=1$. With the help of (\ref{1}), it is easy to
calculate the minimal energy $E_2$ of a pair with a result:
\begin{equation} \label{3} E_2/2 = -3\lambda_S/8\end{equation}
(here the energy per a particle is indicated).

Now consider a cluster of three particles.  The state with
$S=3/2,\ Q=3/2$ has the maximum energy and is of no interest for
us. Passing to consideration of the states with $S=3/2,\ Q=1/2$ or
$S=1/2,\ Q=3/2$ with the same energy $E_3'$, we obtain:
\begin{equation} \label{4} E_3'/3=-7\lambda_S/12\end{equation}
(per a particle as well). These states are doubly degenerated for
the fixed quantum numbers $S,\ S_3,\ Q,\ Q_3$. At last, for the
energy $E_3''$ of a state with $S=1/2,\ Q=1/2$ we have:
\begin{equation} \label{5}E_3''/3=-3\lambda_S/4\ .\end{equation}
This state  is not degenerated for the fixed quantum numbers $S,\
S_3,\ Q,\ Q_3$.

{\bf Model}. Here we restrict ourselves to the simplest model. For
the determination of the equilibrium properties, pinned clusters
and a gas of mobile carriers (fermions) are considered without any
interactions. An interaction of fermions with clusters is taken
into account only for estimation of resistivity. The energy of the
system is written as:
\begin{eqnarray}\nonumber E=\sum_{{\bf
p},\sigma}\Bigl[\epsilon({\bf p})+H\sigma\Bigr]n_\sigma({\bf p})+\\
\label{6} \sum_{\Sigma,\nu}\kappa(\Sigma)\Bigl[\gamma(\Sigma)
E_0(\Sigma,\nu)+H S_3\Bigr]N(\Sigma,\nu)\ .\end{eqnarray} Here the
first sum relates to the fermions (the summations take places over
the momentum ${\bf p}$ and spin projection $\sigma=\pm 1/2$). The
second sum relates to the clusters. The following notation for the
set of quantum numbers is used: $$(S,S_3,Q,Q_3)\rightarrow \Sigma\
.$$  In the second sum, the summations occur over these quantum
numbers and state numbers $\nu$. The value $\gamma(\Sigma)
E_0(\Sigma,\nu)$ is a cluster energy (with the quantum numbers
$\Sigma$ and in the state $\nu$) and $\gamma(\Sigma)$ is number of
particles in the cluster. The constant $\kappa(\Sigma)$ takes into
account additional degeneracy of a level ($\kappa=2$ for the
states with the energy (\ref{4}) and $\kappa=1$ for all other
states). Finally, $$H\equiv g^*\mu_B B \ .$$ Here $B$ is the
magnetic field, $g^*$ is the effective $g$ - factor (supposed to
be the same for the fermions and clusters), $\mu_B$ is the Bohr
magneton.

Further it is meant that only one cluster can be located in the
every state $\nu$. Apparently, the total number of these states is
of the order of the total number of the particles. One would
think, that in a liquid it is necessary to consider a cluster as a
particle with enough big effective mass and characterize it by
momentum. However, in our opinion, it is not unreasonable to
describe the clusters by localized states because of
imperfections.

{\bf Equilibrium properties}. The cluster characteristics can be
obtained with the help of statistical sum $Z(\nu)$ for a
given state: \begin{eqnarray} \nonumber Z(\nu)- 1=\\
\label{7}\sum_\Sigma
\kappa(\Sigma)\exp\biggl\{-\frac{\gamma(\Sigma)\bigl[E_0(\Sigma,\nu)
-\mu\bigr]+H S_3}{T}\biggr\}\ ,\end{eqnarray} where $\mu$ is the
chemical potential of the system.

The fermion distribution function has the usual form:
\begin{equation} \label{8} n_\sigma({\bf p})=\Bigl\{
\frac{\epsilon({\bf p})-\mu_\sigma}{T}+1\Bigr\}^{-1}\ ,\ \ \ \
\mu_\sigma\equiv\mu-H\sigma\ .\end{equation} For quadratic
dependence of the fermion energy on the momentum ($\epsilon({\bf
p})=p^2/(2m^*)$), we have the following expression for the fermion
concentration: \begin{equation} \label{9}n_\sigma =
\frac{1}{V}\sum_{\bf p} n_\sigma({\bf p})=
\frac{m^*T}{2\pi}\ln\Bigl\{1+\exp(\mu_\sigma/T) \Bigr\}
\end{equation} (for a fixed spin projection and for a fixed
valley).

As usual, the chemical potential $\mu$ is defined by the condition
that the total number of the particles is equal to the sum of the
numbers of fermions and electrons inside the clusters. In the case
of two valleys, this condition looks as: \begin{eqnarray}
\nonumber
n= n_F+ n_\Gamma\ ,\ \ \ \  n_F=2\sum_\sigma n_\sigma\ ,\\
\label{10} n_\Gamma=\sum_{\Sigma,\nu} \frac{-T}{Z(\nu)}
\frac{\delta Z(\nu)}{\delta E_0(\Sigma,\nu)}\ .\end{eqnarray} Here
$n$ is the total concentration of electrons, $n_F$ is the
concentration of mobile carriers (fermions), and $n_\Gamma$ is the
concentration of electrons inside the clusters (it is determined
by a functional derivative of the function $Z(\nu)$, see the
expression (\ref{7})).

Up to now, the various expressions were written in general form
that took into account the clusters containing the different
numbers of electrons. Further, we restrict ourselves to clusters
with three electrons for which the more low energies are obtained
than for pairs (see (\ref{3}) - (\ref{5})). For these clusters, it
is necessary to take into account not only the states with lowest
energy (\ref{5}) but also the states with energies (\ref{4})
because, in a sufficiently high magnetic field, a contribution of
clusters with spin $3/2$ is able the main. Furthermore, for
simplicity, we neglect with dispersion of the cluster levels, i.e.
we consider that every energy $E_0$ does not depend on the state
number $\nu$. Instead of the designation $E_0(\Sigma)$ for which
it is necessary all the time to indicate a set of the quantum
numbers, further it is convenient to use some other designations,
namely: \begin{eqnarray}\nonumber
E_0(1/2,1/2) \rightarrow \epsilon_0\ , \ \ \ \ \ \ \ \ \  \\
\label{11} E_0(1/2,3/2)=E_0(3/2,1/2) \rightarrow
\epsilon_1\end{eqnarray} (the arguments correspond to the spin and
quasi-spin values). As appears from the expressions (\ref{4}) and
(\ref{5}), $\epsilon_0<\epsilon_1$.

After these remarks, the statistical sum (\ref{7}) is written in
the form: \begin{eqnarray} \nonumber Z=1+
4\exp\Bigl[-3(\epsilon_0-\mu)/T\Bigr]\cosh\frac{H}{2T}+\\
\label{12} 8\exp\Bigl[-3(\epsilon_1-\mu)/T\Bigr] \biggl\{
\cosh\frac{3H}{2T}+ 3\cosh\frac{H}{2T} \biggr\}\ .\end{eqnarray}
For the concentration of electrons inside the clusters, we have:
\begin{equation} \label{13} n_\Gamma= 3n_0\frac{Z-1}{Z}\
,\end{equation} where $n_0$ is the concentration of the cluster
states.

Now let us find the spin polarization $M$ depending on a magnetic
field. Its ratio to the maximum possible value, $M_m=n/2$, can be
written in the form: \begin{eqnarray} \nonumber \frac{M}{M_m}
=\eta\frac{2T}{3Z}\biggl(\frac{\partial Z}{\partial H} \biggr)_\mu
+\\\label{14}\frac{T}{2\epsilon_F} \Biggl\{
\frac{H}{2T}+\ln\frac{\cosh\bigl[(\mu+H/2)/(2T)\bigr]}
{\cosh\bigl[(\mu-H/2)/(2T)\bigr]} \Biggr\}\ .\end{eqnarray} Here
the first term in the right hand side is the cluster contribution,
and the second term is the fermion contribution. The constant
$\eta$ corresponds to the maximum part of the electrons which can
be located inside the cluster states (apparently, one can think
that $\eta$ is less and of the order of unity), $\epsilon_F$ is
the Fermi energy for the two-valley case at $T=0$ and without
taking into account any clusters: \begin{equation} \label{15}\eta=
\frac{3n_0}{n}\ ,\ \ \ \ \ \epsilon_F= \frac{\pi n}{2m^*}\
.\end{equation}

{\bf Zero temperature}. Let $\epsilon_0<\epsilon_F$. In a magnetic
field and at $T=0$, it is necessary to take into accounts only the
lowest levels of the clusters, namely: $$\epsilon_0\rightarrow
\epsilon_0-(H/2)/3\ ,\ \ \ \ \epsilon_1\rightarrow \epsilon_1-H/2\
.$$ The energy shifts for electrons are
$\delta\epsilon_{\uparrow,\downarrow} =\pm H/2$. As a result, we
have the following expressions for concentrations of fermions with
the different spin projections and electrons inside the clusters
(with minimal energy):
\begin{eqnarray}
\nonumber n_\uparrow =\frac{m^*}{\pi}(\epsilon_0- 2H/3)\ , \\
n_\downarrow = \frac{m^*}{\pi}(\epsilon_0+ H/3)\ ,\label{16}\\
\nonumber n_\Gamma=n- \frac{m^*}{\pi}(2\epsilon_0- H/3)\ ;\\
\nonumber \Bigl(0<H<3\epsilon_0/2\Bigr)\ .\end{eqnarray} In the
greater magnetic fields, \begin{equation}
\label{17}3\epsilon_0/2<H <3(2\epsilon_F-\epsilon_0)\
,\end{equation} a devastation up to zero of the cluster levels and
an increase of $n_\downarrow\rightarrow n$  take places (according
to the linear laws). These behaviors are without taking into
account the levels with the energy $\epsilon_1$ that is in the
case $\epsilon_1> 2\epsilon_F$.

Now, let $\epsilon_0<\epsilon_1<2\epsilon_F$. The level with
energy $\epsilon_1$ comes into the play at magnetic field
$H/3>\epsilon_1 -\epsilon_0$. There are the two cases. Firstly,
this level can be connected up under $n_\uparrow=0$, i.e. when the
magnetic field falls into the interval (\ref{17}). In this case,
the cluster levels are devastated not completely (at the left
border of the interval (\ref{17}), i.e. under
$\epsilon_1=3\epsilon/2$, these levels are not devastated at all).
Secondly, this level comes into the play under $n_\uparrow>0$,
i.e. under $\epsilon_1<3\epsilon_0/2$. Let us discuss the latter
case.

In the range of magnetic field $0<H<3(\epsilon_1-\epsilon_0)$,
there are the previous expressions (\ref{16}). In the point
$H/3=\epsilon_1-\epsilon_0$, the states with initial energy
$\epsilon_1$  are filled up instead of the states $\epsilon_0$
and, without taking into account a dispersion of the cluster
levels, that takes place by a jump. In the range of magnetic field
$3(\epsilon_1-\epsilon_0)<H<\epsilon_1$, we have: \begin{eqnarray}
\nonumber n_\uparrow =\frac{m^*}{\pi}(\epsilon_1- H)\
,\\ n_\downarrow =\frac{m^*}{\pi}\epsilon_1\ , \label{18}\\
n_\Gamma=n- \frac{m^*}{\pi}(2\epsilon_1-H)\ .\nonumber
\end{eqnarray} That continues up to the magnetic field $H=\epsilon_1$,
i.e. up to the complete polarization, when the value of
$n_\uparrow$ equals to zero and the values of $n_\downarrow$ and
$n_\Gamma$ become saturated. The dependence of any quantity,
including a resistivity, on a magnetic field has a form of a
broken curve, and for a magnetic moment even with jumps. It is
clear that the curves are smoothed by temperature.

We note that the above expressions are valid until the obtained
number of electrons inside the clusters is smaller than a maximal
possible one, i.e., until $n_\Gamma<3n_0$. That takes place under
the condition $$1-\frac{\epsilon_1}{2\epsilon_F}<\eta\ .$$

Now about the magnetic moment. Let us consider the case
$\epsilon_0<\epsilon_F,\ \epsilon_0<\epsilon_1<3\epsilon_0/2$ (see
(\ref{16}), (\ref{18}) for the two intervals of a magnetic field).
We obtain for the low magnetic fields: \begin{eqnarray} \label{19}
\frac{M}{M_m} =\frac{5H}{9\epsilon_F}+\frac{1}{3}
\biggl(1-\frac{\epsilon_0}{\epsilon_F}\biggr)\ ,\\ \nonumber
0<H<3(\epsilon_1-\epsilon_0)\ . \end{eqnarray} For the high
magnetic fields up to the complete polarization, we have:
\begin{eqnarray} \label{20} \frac{M}{M_m} =1 +
\frac{H-\epsilon_1}{\epsilon_F}\ ,\\ \nonumber
3(\epsilon_1-\epsilon_0) <H<\epsilon_1\ .\end{eqnarray} The jumps
of the magnetic moment take places at the points $H=0$ and
$H=3(\epsilon_1-\epsilon_0)$. The ratio of susceptibilities in the
two ranges of low and high magnetic field is $$ \frac{\chi_1}
{\chi_2}=\frac{5}{9}\ .$$  In considered case, the complete
polarization takes place under $H=\epsilon_1<3\epsilon_0/2
<3\epsilon_F/2$, while without the clusters under $H=2\epsilon_F$.

In connection with the jump of magnetic moment near the $H=0$, it
is necessary to note the following. In the work \cite{7}, the
direct measurements of the thermodynamic spin susceptibility were
produced. It is revealed that magnetization behaves by nonlinear
manner, the spin susceptibility becoming greater under magnetic
field lowering. Possibly, that is the effect of the clusters.

{\bf Resistivity}. For the resistivity $\rho$, we use the ordinary
expression: \begin{equation} \label{21} \rho=\frac{m}{n_F
e^2\tau}\ ,\end{equation} where $\tau$ is the relaxation time. The
question is what is necessary to take as the $\tau$? If an
interaction of the mobile carriers with impurities is weak (not so
with the clusters) or the number of the impurities is sufficiently
small, then the relaxation time is proportional to an impurity
concentration, i.e., for the scattering by clusters, $1/\tau \sim
n_\Gamma$. That is the main for the sufficiently pure specimens.
Perhaps, it is necessary to take into account the other scatterers
too, then, in the simplest case, one can write: \begin{equation}
\label{22} 1/\tau\sim n_\Gamma+\alpha\ n_i,\end{equation} where
$n_i$ is the concentration of the other scatterers and the factor
$\alpha$ takes into account a difference of the scatterers. As a
result, taking into account the expressions (\ref{13}) and
(\ref{10}), one can write: \begin{equation} \label{23} \rho\sim
\frac{\eta(Z-1)+A\ Z}{(1-\eta)Z+\eta} \ ,\ \ \ \ \ \
A\equiv\alpha\frac{n_i}{n}\ .\end{equation}

Let us consider the low temperatures $T<<\epsilon_F$. Under $H=0$,
from the expressions (\ref{10}) and (\ref{13}), for determination
of the chemical potential, we have the equation:\begin{equation}
\label{24}n-\frac{2m^*\mu}{\pi}= \frac{3n_0}{(1/4)\exp\Bigl[
3(\epsilon_0-\mu)/T \Bigr]+1}\end{equation} (excluding the cluster
states with the energy $\epsilon_1>\epsilon_0$). If one looks for
the chemical potential in the form $$\mu=\epsilon_0+\mu_1\ ,$$
then one can write the equation (\ref{24}) as follows:
\begin{equation} \label{25} 1-\frac{\epsilon_0}{\epsilon_F}
=\frac{\mu_1}{\epsilon_F}+\frac{\eta}{(1/4)\exp(-3\mu_1/T)+1}\
.\end{equation}

Under a partial filling of the cluster levels (for that it is
necessary $\epsilon_0<\epsilon_F$) and at $T=0$, $\mu=\epsilon_0$.
And that is right also at $T\neq 0$ for certain values of
parameters, namely:
\begin{equation} \label{26}1-\frac{\epsilon_0} {\epsilon_F} =
\frac{4\eta}{5}\ ,\ \ \ \ \ \ \mu_1=0\ .\end{equation} If that is
the case then the cluster number is a constant. In this case, a
resistivity does not change while the expression (\ref{22}) is
valid. So, we have the condition for a separatrix.

For small changes of the parameters, one can ignore the first term
in the right-hand side of the (\ref{25}), so that one obtains
approximately: \begin{equation} \label{27} \frac{3\mu_1}{T}\approx
-\ln\biggl\{5\frac{\beta_R}{\beta_L}-4\biggr\}\ ,\end{equation}
where, for convenience, the designations $\beta_L$ and $\beta_R$
are used for the left-hand and right-hand sides of the (\ref{26}),
accordingly. From the (\ref{27}), one can see that, under the
condition $\beta_R/\beta_L>1$ ($\mu_1<0$), the fermion number
decreases and the number of scatterers increases, therefore the
resistivity increases with temperature. Such behavior corresponds
to a metallic phase. Otherwise, we have an insulator phase. In the
both cases, the resistivity behaves in accordance with a linear
law.

Resistance behavior with temperature depends on a value of a
magnetic field. Let us consider a magnetic field in the interval
given for the expression (\ref{18}). In this case, it is enough to
take into account the cluster state only with the spin $3/2$ and
quasi-spin $1/2$. The previous expressions (\ref{24}) -
(\ref{26}), after the replacement
$$\epsilon_0\rightarrow (\epsilon_1-H/2)\ ,$$ remain valid.
For example, instead of the (\ref{26}) we come to:
\begin{equation} \label{28}1- \frac{\epsilon_1-H_S/2} {\epsilon_F}
= \frac{4\eta}{5}\ ,\ \ \ \ \ \ \mu_1=0\ ,\end{equation} where
$H_S$ is the magnetic field corresponding to a separatrix. In view
of $H_S<\epsilon_1$, we obtain the following condition for origin
of separatrix:
$$\frac{\epsilon_1}{2\epsilon_F} <1-\frac{4\eta}{5}\ .$$

Under deviation from the (\ref{28}), the expression for $\mu_1$
coincides with the (\ref{27}) but, in place of the previous
left-hand and right-hand parts, it is necessary to substitute the
corresponding parts of the (\ref{28}). With magnetic field, the
left-hand part grows whereas the right-hand part is constant.
Then, as follows from the (\ref{27}), for the lower magnetic
field, $H<H_S$, we have the metallic phase (the resistivity grows
with temperature) and for the greater magnetic field, $H>H_S$, we
have the insulator phase (the resistivity decreases with
temperature increasing).

Physical meaning of the appearance of separatrix is simple. When a
filling of the cluster states is small then it increases with
temperature. Otherwise, for a large filling, it decreases with
temperature increasing. The cluster filling grows with the
magnetic field. At some magnetic field, an intermediate case can
take place and the cluster filling is constant. So, a separatrix
appears. Thus, when one draws a conclusion from the temperature
dependance of resistivity then the metal - insulator transition
can happen. Such a behavior of resistivity corresponds
qualitatively to the experiments.

In the figure, the resistivity dependence on magnetic field is
shown for different temperatures, namely: $T/\epsilon_F=0.05$
(dots), $T/\epsilon_F=0.1$ (solid line), and $T/\epsilon_F=0.15$
(dashed line). The following parameters are taken:
$\epsilon_0/\epsilon_F=0.95$, $\epsilon_1/\epsilon_F=1.1$,
$\eta=0$, and $A=0$. The $\rho_0$ is resistivity under $T=H=0$.
The point of intersection of the curves corresponds to separatrix.

 The conclusion about enhancement of the resistivity under influence
 of the magnetic field is the direct consequence of our model accounting
 for the presence of the clusters. The validity of the expressions
 (\ref{21}) and (\ref{22}) is assumed. That would be the case if the
 interaction of the mobile carriers with the clusters could be considered as weak.
 For strong interaction, the expression (\ref{22}) can be used only
 in the limit of small number of the clusters, $n_\Gamma<<n$,
 possibly up to $n_\Gamma\sim n$. However, for some values of the
 $n_\Gamma$ (of the order of $n$), that simple picture can be incorrect
 because it is necessary to take into account the more complicated
 phenomena such as, e.g., localization of the fermions.

 I am grateful to A.V.Chaplik, M.V.Entin, Z.D.Kvon and G.I.Surdutovich
 for helpful discussions. This work was supported by the Russian Foundation
 for Basic Research (project no. 02-02-16159) and INTAS (grant no.
 2212).

\newpage
\begin{figure}[ht]
\begin{center}
\centerline{\input epsf \epsfysize=8 cm \epsfbox{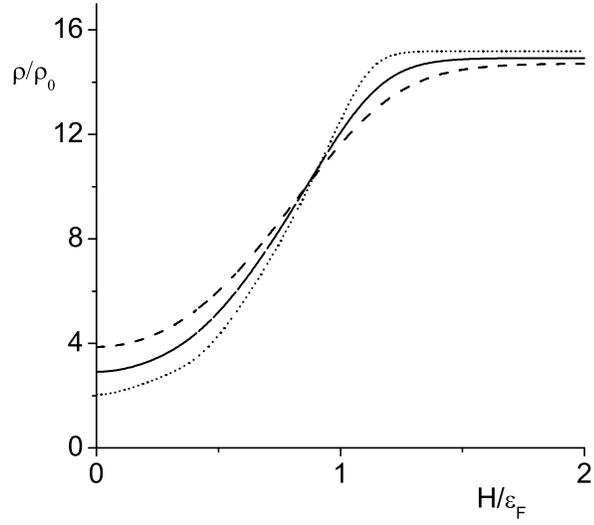}}
\end{center}
\caption{Resistivity dependence on magnetic field for different
temperatures (see text)}
\end{figure}

\end{document}